\documentclass{article}
\usepackage[utf8]{inputenc}
\usepackage{epsfig,latexsym,amsfonts,amsmath,amsthm,amssymb,amsbsy,multirow,slashed,wasysym,textcomp,comment,mathtools,bbold,cite,datetime}
\usepackage{geometry}
 \usepackage{pgfplots}
 \usepgfplotslibrary{patchplots}
 \pgfplotsset{compat=newest}
 
\usepackage{hyperref}
 \hypersetup{
      linktoc=all,    
  }

\usepackage{blindtext}

\usepackage{tikz}
\usetikzlibrary{arrows.meta}
\tikzstyle{bag} = [align=center]
\usetikzlibrary{decorations.pathmorphing}

\def\bz{{\bar z}}
\def\bw{{\bar w}}

 \newcommand{\badat}{\begin{alignedat}}
 \newcommand{\eadat}{\end{alignedat}}

\theoremstyle{definition}

\DeclareGraphicsExtensions{.pdf,.png}
\usetikzlibrary{patterns}
\usetikzlibrary{math}

\usetikzlibrary{arrows.meta}
\tikzstyle{bag} = [align=center]
\usetikzlibrary{decorations.pathmorphing}
\usepackage[normalem]{ulem}

\usetikzlibrary{calc}

\def\aaa{\alpha}
\def\bbb{\beta}

\begin{document}
 \begin{titlepage}
 \begin{flushright}{${}$}
 \end{flushright}
	 \vspace{1.5cm}
\begin{center}
  \baselineskip=13pt 
    {\huge  Mapping SYK to the Sky}
   \vskip1.5cm 
   {\large Sabrina Pasterski}${}^\diamondsuit$
   {\large and Herman Verlinde}${}^\blacklozenge$
  \\[8mm]
   {\em${}^\blacklozenge$ Physics Department, Joseph Henry Laboratories}\\[3mm]
\noindent{${}^\diamondsuit$ \it Princeton Center for Theoretical Science}\\[3mm]
{\em Princeton University, Princeton, NJ 08544, USA}

\end{center}

\vspace{2cm}

\begin{abstract}
\addtolength{\baselineskip}{.5mm}

The infrared behavior of gravity in 4D asymptotically flat spacetime exhibits a rich set of symmetries.  This has led to a proposed holographic duality between the gravitational $\mathcal{S}$-matrix and a dual field theory living on the celestial sphere. Most of our current understanding of the dictionary relies on knowledge of the 4D bulk.  As such, identifying intrinsic 2D models that capture the correct symmetries and soft dynamics of 4D gravity is an active area of interest.  Here we propose that a 2D generalization of SYK provides an instructive toy model for the soft limit of the gravitational sector in 4D asymptotically flat spacetime. We find that the symmetries and soft dynamics of the 2D SYK model capture the salient features of the celestial theory: exhibiting chaotic dynamics, conformal invariance, and a $w_{1+\infty}$ symmetry. The holographic map from 2D SYK operators to the 4D bulk employs the Penrose twistor transform.

\end{abstract}

\vfill

\end{titlepage}


\addtolength\baselineskip{1mm}
\tableofcontents

\setcounter{tocdepth}{3}

\addtolength\parskip{.5mm}
\addtolength\baselineskip{-.5mm}
\def\is{\! & \! = \! & \!}
\def\half{{1\over 2}}
\numberwithin{equation}{section}
\def\sub{\scriptscriptstyle}
\def\ip{${\mathcal I}^+$}
\def\calr{{\cal R}}
\def\e{{\epsilon}}
\def\g{{\gamma}}
\def\cs{{\cal S}}
\def\S{\Sigma }
\def\s{\sigma }
\def\sz{\sigma^{0} }
\def\Psz{\Psi^{0} }
 \def\p{\partial}
 \def\bz{{\bar z}}
 \def\cT{8\pi G_N {\mathcal T}}
\def\0{{(0)}}
\def\1{{(1)}}
\def\2{{(2)}}
 \def\cL{{\cal L}}
\def\co{{\cal O}}\def\cv{{\cal V}}
\def\n{\nabla}
\def\ci{{\mathcal I}}
\def\ipp{${\mathcal I}^+_+$}
\def\<{\langle }
\def\>{\rangle }
\def\[{\left[}
\def\]{\right]}
\def\bw{{\bar w}}
\def\pphi{\phi}
\def\h{{h^i_i}}
\def\t{{\rm{trace}}}
\def\o{\omega }
\def\ra{\bigr\rangle}
\newcommand{\non}{\nonumber}
\renewcommand{\O}{\Omega}
\renewcommand{\L}{\Lambda}
\newcommand{\bigO}{\mathcal{O}}
\newcommand{\sech}{\mbox{sech}}
\newcommand{\tn}{\tilde{n}}
\newcommand{\W}{\mathcal{W}}
\newcommand{\tr}{\mbox{tr}}
\newcommand{\Del}{\nabla}
\newcommand{\hs}[1]{\mbox{hs$[#1]$}}
\newcommand{\w}[1]{\mbox{$\W_\infty[#1]$}}
\newcommand{\bif}[2]{\small\left(\!\!\begin{array}{c}#1 \\#2\end{array}\!\!\right)}
\renewcommand{\u}{\mathfrak{u}}
\newcommand{\scriplus}{\mathcal{I}^+}
\renewcommand{\epsilon}{\varepsilon}
\def\dim#1{\lbrack\!\lbrack #1 \rbrack\!\rbrack }
\newcommand{\chichi}{\chi\!\cdot\!\chi}
\newcommand\snote[1]{\textcolor{magenta}{[S:\,#1]}}

\renewcommand{\theequation}{\thesection.\arabic{equation}}
   \makeatletter
  \let\over=\@@over \let\overwithdelims=\@@overwithdelims
  \let\atop=\@@atop \let\atopwithdelims=\@@atopwithdelims
  \let\above=\@@above \let\abovewithdelims=\@@abovewithdelims
\renewcommand\section{\@startsection {section}{1}{\z@}%
                                   {-3.5ex \@plus -1ex \@minus -.2ex}
                                   {2.3ex \@plus.2ex}%
                                   {\normalfont\large\bfseries}}

\renewcommand\subsection{\@startsection{subsection}{2}{\z@}%
                                     {-3.25ex\@plus -1ex \@minus -.2ex}%
                                     {1.5ex \@plus .2ex}%
                                     {\normalfont\bfseries}}

\newcommand{\Tr}{\mbox{Tr}}
\renewcommand{\H}{\mathcal{H}}
\newcommand{\SU}{\mbox{SU}}
\newcommand{\chiu}{\chi^{{\rm U}(\infty)}}
\newcommand{\ff}{\rm f}
\linespread{1.3}

\def\bfR{{\mbox{\textbf R}}}
\def\gzz{\gamma_{z\bz}}
\def\vx{{\vec x}}
\def\p{\partial}
\def\po{$\cal P_O$}
\def\cN{{\cal H}^+ }
\def\N{${\cal H}^+  ~~$}
\def\G{\Gamma}
\def\l{{\ell}}
\def\ch{{\cal H}^+}
\def\Q{{\hat Q}}
\def\T{\hat T}
\def\C{\hat C}
\def\zet{z}
\def\scc{\mbox{\small $\hat{C}$}}
\def\Aa{{\mbox{\scriptsize \smpc \sc a}}}
\def\aalpha{{\mbox{\scriptsize {\smpc  $\alpha$}}}}
\def\cC{{\mbox{\scriptsize {\smpc \sc c}}}}
\def\cCb{{\mbox{\scriptsize \smpc \sc c'}}}
\def\sS{{\mbox{\scriptsize {\smpc \sc s}}}}
\def\Bb{{\mbox{\scriptsize \smpc \sc b}}}
\def\Hh{{\mbox{\scriptsize \smpc \sc h}}}
\def\oO{{}} 
\def\bfC{\mbox{{\textbf C}}}
\def\nonu{\nonumber}
\def\im{{\rm i}}
\def\tr{{\rm tr}}
\def\be{\bea}
\def\ee{\eea}

\def\llambda{{z}}
\def\spc{\hspace{.5pt}}

\def\bea{\begin{eqnarray}}
\def\eea{\end{eqnarray}}
\def\half{{\textstyle{\frac 12}}}
\def\cL{{\cal L}}
\def\halfi{{\textstyle{\frac i 2}}}

\def\delbar{\overline{\partial}}
\newcommand{\smpc}{\hspace{.5pt}}
\def\nspc{\!\spc\smpc}
\def\uU{\mbox{\textit{\textbf{U}}\spc}}
\def\uV{\mbox{\textit{\textbf{U}\spc}}}
\def\tT{\mbox{\textit{\textbf{T}}\spc}}
\def\bfC{\mbox{\textit{\textbf{C}}}}
\def\bn{\mbox{\textit{\textbf{n}\!\,}}}
\def\pP{\mbox{\textit{\textbf{P}\!\,}}}
\def\rR{{\textit{\textbf{R}\!\,}}}

\def\im{{\rm i}}
\def\tr{{\rm tr}}

\def\ra{\bigr\rangle}
\def\la{\bigl\langle}
\def\li{\bigl |\spc}
\def\ri{\bigr |\spc}

\def\nonu{\nonumber}
\def\SL2{SL(2,\mathbb{R})}
\def\mR{\mathbb{R}}
\def\mZ{\mathbb{Z}}
\def\nn{\nonumber}

\def\centerarc[#1](#2)(#3:#4:#5)
    { \draw[#1] ($(#2)+({#5*cos(#3)},{#5*sin(#3)})$) arc (#3:#4:#5); }

\enlargethispage{\baselineskip}

\setcounter{tocdepth}{2}
\addtolength{\baselineskip}{.3mm}
\addtolength{\parskip}{.3mm}
\addtolength{\abovedisplayskip}{.9mm}
\addtolength{\belowdisplayskip}{.9mm}
\renewcommand\Large{\fontsize{15.5}{16}\selectfont}

\newcommand{\newsubsection}[1]{
\vspace{.6cm}
\pagebreak[3]
\addtocounter{subsubsection}{1}
 \addcontentsline{toc}{subsection}{\protect
 \numberline{\arabic{section}.\arabic{subsection}.\arabic{subsubsection}}{#1}}
\noindent{\arabic{subsubsection}. \bf #1}
\nopagebreak
\vspace{1mm}
\nopagebreak}
\renewcommand{\footnotesize}{\small}
\def\mmu{\zeta}
\def\biz{{\tilde{z}}}

\section{Introduction}
\vspace{-2mm}

Celestial holography proposes a duality between gravity in asymptotically flat spacetimes and a conformal field theory living on the celestial sphere~\cite{Strominger:2017zoo,Raclariu:2021zjz,Pasterski:2021rjz}.  This presentation of the gravitational $\mathcal{S}$-matrix has shed light on the infinite dimensional symmetry enhancements that arise in the infrared limit of scattering~\cite{Strominger:2013jfa}.  Indeed the starting point of this program is the observation that soft theorems in 4D look like currents in 2D when we transform our external particles from momentum to boost eigenstates. 

So far, most of our understanding of celestial holography is derived from our knowledge of the 4D physics and how it maps to 2D under our change of basis~\cite{Pasterski:2016qvg,Pasterski:2017kqt,Pasterski:2017ylz}.  While this change of basis depends on physics at all energy scales, we can make robust statements about the holographic dictionary that are agnostic about the UV completion by focusing on aspects  that are governed by symmetries.   The asymptotic symmetry group of gravity in the bulk is infinite dimensional and any choice of vacuum breaks all but a finite number of the symmetry generators.  The so-called `conformally soft sector' of scattering describes the dynamics of associated Goldstone modes~\cite{Cheung:2016iub,Donnay:2018neh,Arkani-Hamed:2020gyp,Pasterski:2021dqe}.

Studying the Goldstone mode dynamics can provide a surprising amount of mileage. First, this sector 
provides non-trivial constraints on scattering. In going to the boost basis, the soft modes that generate the asymptotic symmetries are recast as celestial currents that couple to matter via conformally soft theorems~\cite{Puhm:2019zbl,Adamo:2019ipt,Guevara:2019ypd}.  The corresponding celestial OPEs give rise to differential constraints on amplitudes and provide a first step towards a celestial bootstrap program. Second, this is the perfect arena for developing candidate boundary dual models that capture sectors of the bulk theory~\cite{Cheung:2016iub, Nguyen:2020hot}.  Exploring these models informs our understanding of the relevant representations of the asymptotic symmetry group.

In the recent paper~\cite{Pasterski:2022lsl}, we applied both of these approaches to the celestial Virasoro symmetry, identifying a sector with large central charge that incorporates backreaction effects and exhibits maximal quantum chaos.  By focusing on the connection between radial evolution in the celestial CFT and Rindler evolution in the bulk we were able to observe Lyapunov behavior without needing to add horizons in the bulk -- the observer horizon of the Rindler trajectories taking its place.   Our emphasis on the Virasoro multiplets also avoided a discussion of how translation invariance and other symmetries manifest in this story, which we would like to remedy. While this is convenient from the perspective of using 2D CFT machinery, part of the richness of the celestial dual is that gravity in 4D asymptotically flat spacetimes possesses a much larger set of symmetries. In particular, there exists a semi-infinite tower of conformally soft theorems which exhibit a $w_{1+\infty}$ symmetry~\cite{Guevara:2021abz,Strominger:2021lvk} that has received a flurry of attention recently~\cite{Himwich:2021dau,Adamo:2021lrv,Ball:2021tmb,Mago:2021wje,Freidel:2021ytz,Costello:2022wso}.

We are led to the following question: can we build a model that incorporates the full set of infinite dimensional symmetry enhancements associated to 4D gravity and gives an origin for the signals of chaos we saw in~\cite{Pasterski:2022lsl}? The goal of this paper is to answer in the affirmative. We show that a 2D generalization of the Sachdev-Ye-Kitaev (SYK) model closely related to the one proposed in~\cite{Turiaci:2017zwd} provides an example of a quantum many body system with soft dynamics that strongly resembles, or in an appropriate realization even matches, the conformally soft sector of gravity in 4D. Like its 1D archetype~\cite{KitaevTalk2v2,SY1993,Sachdev:2015efa,Polchinski:2016xgd,Maldacena:2016hyu}, this 2D generalization is known to demonstrate maximally chaotic behavior, exhibit conformal symmetry in the IR, and be invariant under area preserving diffeomorphisms -- three features our desired dual must possess!

This paper is organized as follows. In section~\ref{sec:celestial_w} we review the celestial incarnation of the $Lw_{1+\infty}$ symmetry: as manifested in analytic continuations of the collinear limits of positive helicity gravitons and its twistor description in self-dual gravity.  We then review a 2D generalization of the SYK model based on~\cite{Turiaci:2017zwd} in section~\ref{sec:2DSYK}, emphasizing the symmetries exhibited by this model in the infrared, which include a Schwarzian Goldstone mode and area preserving diffeomorphisms. Finally, we connect these two stories in section in section~\ref{sec:toy_dual}, showing how the soft dynamics of the 2D model of~\cite{Turiaci:2017zwd} captures the main features of the conformally soft sector of gravity in 4D asymptotically flat spacetime.

\section{Celestial \texorpdfstring{$\boldsymbol{w_{1+\infty}}$}{w-infinity}}\label{sec:celestial_w}

\vspace{-1mm}

The celestial holographic map requires us to change our external scattering states from energy to boost eigenstates~\cite{Pasterski:2016qvg,Pasterski:2017kqt}.  For massless scattering this can be achieved by exchanging lightcone energy $\omega$ 
\bea
\label{ptosphere}
p^\mu = \pm\omega q^\mu , \qquad q^\mu \! \is \! \frac1 2  \bigl( 1\nspc +\nspc  \biz z, z \nspc +\nspc  \biz, i (\biz\nspc - \nspc z), 1\nspc -\nspc z\biz \bigr),
\eea
for a 4D  Rindler energy, equivalent to a 2D conformal weight $\Delta$, via a Mellin transform
\bea\label{Mellin}
A(\Delta_i, z_i,\biz_i) \is \Bigl[ \, \prod_i \int_0^\infty\!\! d\omega_i\, \omega_i^{\Delta_i-1} \Bigr] 
A(\omega_i;z_i,\biz_i)\,.
\eea
The resulting object transforms like a correlator of $\mathrm{SL}(2,\mathbb{C})$ primaries $\mathcal{O}^\pm_{\Delta,J}(z,\biz)$ with conformal dimension $\Delta$ and spin $J=\ell$ (matching the 4D helicity) at the location on the celestial sphere corresponding to the direction of the null momentum~\eqref{ptosphere}.  Unless necessary, we will suppress the superscript indicating $in$ and $out$. We will also go freely between (1,3) and (2,2) signatures, in which case it is more convenient to use the left and right conformal weights $(h,\bar{h}) =(\frac{1}{2}(\Delta+J),\frac{1}{2}(\Delta-J)).$ While the principal series spectrum $\Delta=1+i\lambda$ provides a basis of finite energy states~\cite{Pasterski:2017kqt}, translations generate a shift in the spectrum $\Delta\rightarrow\Delta+1$~\cite{Donnay:2018neh,Stieberger:2018onx} and we will want to analytically continue to generic complex $\Delta\in\mathbb{C}$ in what follows~\cite{Donnay:2020guq}.

  \def\ww{{\hat w}}

\subsubsection*{$w_{1+\infty}$ from collinear limits of scattering}

We start with a brief review of how the $w_{1+\infty}$ symmetry arises in this basis, following~\cite{Strominger:2021lvk}.  Consider the sector of positive helicity gravitons. As discussed above, the structure of celestial CFT leads us to consider operators with conformal dimensions analytically continued to the complex plane $\Delta\in\mathbb{C}$.  If we look at a soft expansion of the momentum space amplitude, we see that subleading powers in $\omega\rightarrow0$ turn into residues at integer values of the conformal dimension, namely (at tree level)~\cite{Pate:2019mfs}
 \be\label{alim}
\lim\limits_{\Delta\rightarrow-n}(\Delta+n)\int_0^\infty d\omega \omega^{\Delta-1}\sum_k\omega^k A^{(k)}=A^{(n)}.
\ee
When the external leg is a graviton, the expansion starts with the leading soft theorem~\cite{Weinberg:1965nx} at $k=-1$ corresponding to the Ward identity for supertranslations~\cite{He:2014laa}, followed by the subleading soft graviton~\cite{Cachazo:2014fwa} at $k=0$ corresponding to superrotations~\cite{Kapec:2014opa}.

With this interpretation in mind, let us define the corresponding residues as follows
 \bea\label{hk}
H^k\! \is \! \lim\limits_{\epsilon\rightarrow0} \epsilon\, \mathcal{O}_{k+\epsilon,+2}~~~k=2,1,0,-1...
 \eea
 where $\mathcal{O}_{k+\epsilon,+2}$ denotes the positive helicity graviton of conformal weight $(h,\bar{h}) = (\frac{k+2}{2},\frac{k-2}{2})$. 
  This tower of conformal dimensions is special from a purely representation theoretic point of view.  Namely, these each have primary descendants.  For the standard BPZ inner product, primary descendants are null and the corresponding $\overline{\mathrm{SL}(2,\mathbb{C})}$ multiplets are finite dimensional
\be
\Delta=3-n,~~n\in\mathbb{Z}_{>}~~~~\Rightarrow~~~~\bar{L}_{-1}^n|\Delta,J=+2\rangle=0. 
\ee 
While this truncation is not guaranteed from the Hermiticity conditions obeyed by the Lorentz generators with the standard 4D Hilbert space inner product, these descendants are observed to reduce to contact terms at sources for the celestial currents in scattering amplitudes.  
Taking this truncation as an input leads us to the following mode expansion for these residues
 \be\label{trunc}
\frac{1}{\kappa}  H^{-2p+4}(z,\bz)=\sum_{n={1-p}}^{p-1} \, \frac{{\biz}^{p-n-1}}{\! \Gamma(p-n)\Gamma(p+n)\!} \;  W_n^p(z). 
 \ee
 Here we included a prefactor in the expansion, anticipating the identification of $W_n^p(z)$ with holomorphic $w_{1+\infty}$ currents, with corresponding global charges
\bea
\label{wcharges}
\ww^p_n = \oint \! \frac{dz}{2\pi i} \, W^p_n(z).
\eea

 Celestial OPEs are determined by the collinear limits of scattering. 
  The standard procedure is to compute the complexified collinear limits starting from known amplitudes in $\mathbb{R}^{1,3}$ and analytically continue to Klein space where the global Lorentz symmetry becomes $\mathrm{SL}(2,\mathbb{R})\times\mathrm{SL}(2,\mathbb{R})$ and $z$ and $\bz$ are independent real variables (which are then complexified to $\mathbb{C}^2$). 
  Using this procedure, one finds that the positive helicity sector exhibits a rich symmetry algebra~\cite{Guevara:2021abz}.

  The symmetry algebra of the conserved charges are computed using the radial quantization prescription applied to complexified celestial sphere coordinates.  For instance, in the positive helicity sector each of the modes $W^k_n(z)$ would be treated as holomorphic operators.  Via the usual prescription
   \bea\label{radcom}
[A,B](z)\!\is \! \oint_z\frac{dx}{2\pi i}A(x)B(z),
 \eea
one finds that the charges \eqref{wcharges} satisfy the $w_{1+\infty}$ commutation relations~\cite{Strominger:2021lvk}
\bea\label{winfty}
[\ww_m^p,\ww^q_n]\! \is \! [m(q-1)-n(p-1)]\ww_{m+n}^{p+q-2},
\eea
where\footnote{Note that in order to have sources for the celestial currents we need to allow for a contact term at the level $n$ descendants. This is consistent with the truncated $\overline{\mathrm{SL}}(2,\mathbb{C})$ multiplets of~\eqref{trunc} so long as the $z$-dependence has poles.  }
$p=1,\frac{3}{2},2,\frac{5}{2},...$ and $1-p\le m\le p-1.$

The restriction in the range of $p$ takes us to the wedge subalgebra of $w_{1+\infty}$.
Applying the bracket~\eqref{radcom}  to the $z$-dependent modes $W^p_n$, we find a realization of the corresponding subalgebra of the loop algebra $Lw_{1+\infty}$.

\subsubsection*{$w_{1+\infty}$ symmetry of self-dual gravity}\label{sec:sdg}
 To highlight the geometric origin of the celestial $w_{1+\infty}$ symmetry in self-dual gravity, we briefly recall the twistor-space construction of~\cite{Adamo:2021lrv}. 
The connection to self-dual gravity opens us up a rich branch of the literature. The realization of a $w_{1+\infty}$ symmetry algebra within self-dual gravity dates back to to~\cite{Boyer:1985aj,Park:1989fz,Park:1989vq} and was already implicit in the original work by Penrose on the non-linear graviton construction of classical solutions to self-dual gravity~\cite{Penrose:1976js}.

Let $ Z^A=(\mmu^{\dot \aaa},\llambda_\aaa) $ be homogeneous coordinates on $\mathbb{CP}^3$, with twistor space $\mathbb{PT}$ the locus where $\llambda_\alpha\neq 0$. We can write $\mathbb{PT}$  as a fibration over the celestial sphere coordinate $\llambda\in\mathbb{CP}^1$
\be\label{fibra}
p: \mathbb{PT}\rightarrow \mathbb{CP}^1,~~~p(Z)=\llambda_\aaa.
\ee
Points in Minkowski space correspond to a linear embedding of the Riemann sphere into $\mathbb{PT}$ defined by the incidence relation $\mmu^{\dot \aaa}=x^{a \dot \aaa}\llambda_\aaa$.
The $w_{1+\infty}$ symmetry acts via  area preserving diffeomorphisms of the 2-plane parametrized by $\mmu^{\dot \aaa}=(\mmu^{\dot0},\mmu^{\dot1})$ 
\be\label{pst}
\{f,g\}=\epsilon^{\dot \aaa \dot \bbb }\p_{\dot \aaa}f\p_{\dot \bbb} g,~~~\epsilon^{\dot 0 \dot 1}=1.
\ee
Performing a double Taylor series expansion in these coordinates, we can use the following monomials 
\be\label{wmp}
v_{m}^p=\frac{1}{2}(\mmu^{\dot0})^{p+m-1}(\mmu^{\dot1})^{p-m-1},~~~~~~~~~2p-2\in \mathbb{Z}_{\ge},~~ 1-p\le m\le p-1,
\ee
as a basis for functions on the $\xi$-plane. 
The Poisson bracket of these modes
\be
\{v_{m}^p,v_{n}^q\}=(m(q-1)-n(p-1))v_{n+m}^{p+q-2}
\ee
matches the wedge algebra of $w_{1+\infty}$.  In this context, the restriction to the wedge subalgebra comes from demanding that our vector fields are regular at the origin of the $\zeta$-plane.

 The non-linear graviton construction follows from a theorem of Penrose~\cite{Penrose:1976js} that establishes a bijective correspondence between self-dual Ricci flat metrics on regions of  complexified Minkowski space and complex deformations of twistor space that preserve the fibration~\eqref{fibra} and Poisson structure~\eqref{pst} on the fibers. If we cover the celestial sphere with two patches $U=\{\llambda_0\neq 0\}$ and $\widetilde{U}=\{\llambda_1\neq 0\}$ parametrized by $ Z^A=(\mmu^{\dot \aaa},\llambda_\aaa) $ and $\tilde{Z}^A=(\tilde\mmu^{\dot \aaa},\llambda_\aaa) $, turning on the non-linear graviton background amounts to gluing the two patches together via a non-trivial area preserving transition function. 
Infinitesimal deformations of this type that respect the scaling symmetry of $\mathbb{CP}^3$ are determined by a degree two generating function $G(\llambda_\alpha,\mmu^{\dot0},\tilde{\mmu}^{\dot1})$ that prescribe how the 2-plane fibers are patched over the intersection via the canonical transformation
\be
\mmu^{\dot1}=\frac{\p G}{\p\mmu^{\dot0}},~~~\tilde{\mmu}^{\dot0}=\frac{\p G}{\p\tilde{\mmu}^{\dot1}}.
\ee
A basis of such functions takes the form
\be
\label{gpmr}
g^p_{m,r}=\frac{v^p_m}{\llambda_0^{2p-4-r}\llambda_1^r},~~~~~~~~~r\in\mathbb{Z},
\ee
with $\mmu^{\dot1}\mapsto\tilde{\mmu}^{\dot1}$ in~\eqref{wmp}. This is precisely the loop algebra of the wedge algebra of $w_{1+\infty}$.   

By Penrose's theorem these are isomorphic to self-dual Ricci flat solutions.  The conformal primary wavefunctions with positive helicity are exact (complex) self-dual spacetimes of this type~\cite{Pasterski:2020pdk}.   The authors of~\cite{Adamo:2021lrv} explore this $w_{1+\infty}$ symmetry in the context of twistor sigma models sigma models~\cite{Adamo:2021bej}. We will return to the connection between these area preserving diffeomorphisms and the celestial dictionary in the context of our 2D SYK model in section~\ref{sec:toy_dual}.

\section{A 2D generalization of SYK}\label{sec:2DSYK}

We now turn to discuss a 2D quantum many body model that exhibits both the $w_{1+\infty}$ symmetry and the type of chaotic Goldstone mode dynamics we expect for gravitational scattering in asymptotically flat spacetimes. The Sachdev-Ye-Kitaev (SYK) model is of interest as an example of a soluble quantum many body system that exhibits maximal chaotic behavior, and as a candidate holographic dual for 2D black hole spacetimes~\cite{KitaevTalk2v2,SY1993,Sachdev:2015efa,Polchinski:2016xgd,Maldacena:2016hyu}. The standard 1D version of the model consists of $N$ Majorana fermions interacting via a homogeneous non-linear potential with Gaussian random couplings. It exhibits an emergent conformal symmetry in the IR that is modeled by the same Schwarzian theory found in 2D gravity~\cite{Jensen:2016pah,Maldacena:2016upp,Engelsoy:2016xyb}.

2D generalizations of the SYK model were proposed in~\cite{Turiaci:2017zwd} and \cite{Lian:2019axs}.  Here we consider a slight modification of the model of~\cite{Turiaci:2017zwd} with an action of the form
\bea\label{fermi}
& & \qquad\qquad\qquad S\, =\, S_{\rm\smpc UV}+S_{\rm\spc IR},\\[3.5mm]
\label{moduv}
& &\quad S_{\rm \smpc UV} \spc = \, \sum_i \int d^2 z\; \epsilon^{\mu\nu}\psi^i_+\p_\mu \psi^i_+\, \psi_-^i\p_\nu\psi^i_-, \label{suv} \\[1mm]
\label{sir}
S_{\rm\spc IR}\is\!\! \! \int\! d^2 z\Bigl(\spc \sum_{i_1,...,i_q} J^-_{i_1...i_q}\psi^{i_1}_-...\,\psi^{i_q}_-\spc +\! \sum_{j_1,...,j_q} J^+_{j_1...j_q} \psi^{j_1}_+...\,\psi^{j_q}_+\Bigr).
\eea
Here $\psi_\pm^i$ where $i=1,...,N$ are the chiral components of $N$ 2D Majorana fermions, while $J^\pm_{i_1...j_q}$ are Gaussian random couplings with variance
\bea
\bigl\langle (J^\pm_{i_1...j_q})^2 \bigr\rangle\! \is \! \frac{J^2(q-1)!}{N^{q-1}}.
\eea
We briefly compare the symmetries of the UV and IR limits of this theory, in turn. 
The  modified kinetic term  $S_{\rm\smpc UV}$ is designed so that the fermions have canonical scale dimension $[\psi]_{\rm \smpc UV}=0$. This makes the interaction term $S_{\rm \spc IR}$ relevant so that it dominates in the IR. Moreover, for the version of the model considered in \cite{Turiaci:2017zwd}, it was found that the IR Schwinger-Dyson equation exhibits conformal symmetry and that the emergent pseudo-Goldstone action arising from broken reparameterization invariance matches the boundary action derived from 3D AdS gravity. This latter match hints at a potential application to celestial holography, given the analysis in our recent paper \cite{Pasterski:2022lsl}. 

\subsubsection*{UV symmetries and commutators}
\vspace{-1mm}

Let us start by looking at the UV action. It splits into $N$ decoupled terms, consisting of a pair of  Majorana fermions. We will focus on a single copy here.  
To understand the canonical structure of this kinetic term, it is useful to rewrite~\eqref{moduv} by introducing the Hubbard-Stratonovich variables $e_\mu^\pm$ 
\be\label{s_uv2}
S'_{\rm \smpc UV}=\frac{1}{2}\int d^2 z\, \epsilon^{\mu\nu}\Bigl(e^a_\mu\psi_a\p_\nu\psi_a-\frac{1}{2}\epsilon_{ab}e^a_\mu e^b_\nu\Bigr)
\ee
where $a=\pm$. For fixed $e_\mu^\pm$, treated as a Cartan zweibein,  we recognize a standard fermionic kinetic term. We get to~\eqref{moduv} by integrating out the $e_\mu^\pm$. The classical equations of motion we get from varying the action with respect to the zweibein equate $e^+_\mu =-\psi_-\p_\mu\psi_-$ and  $ e^-_\mu =\psi_+\p_\mu\psi_+$. For fixed frame field, the equal time anti-commutation relations read
\bea
\{\psi_\pm(z_1),\psi_\pm(z_2)\}\! \is \! ({e_1^\pm})^{-1}\delta(z_{12}).
\eea
Upon performing a Legendre transform, we get the following free Hamiltonian density
\be\label{h01}
H'_{UV}=e^a_0(\epsilon_{ab}e_1^b-\psi_a\p_1\psi_a).
\ee
We see that $e^a_0$ acts as a Lagrange multiplier. This UV theory is invariant under a local internal Lorentz symmetry 
as well as reparameterizations  
\bea
e^\pm_\mu(z)\; \mapsto\;  \Lambda^{\mp 2}(z) e^\pm_\mu(z),~~& &~~\psi_\pm(z)\; \mapsto\;  \Lambda^{\pm1}(z) \psi_\pm(z)\\[2mm]
\label{reparam}
e^\pm_\mu(z)\; \mapsto\; \frac{\p \tilde{z}^\nu}{\p z^\mu}\, e^\pm_\nu(\tilde{z}(z)),~~& &~~\psi_\pm(z)\; \mapsto\; \psi_\pm(\tilde{z}(z))
\eea
under which the fermions transform like scalars. These will be partially broken by the interaction term, which we turn to next.

\subsubsection*{Symmetries of the interacting theory}

\vspace{-1mm}

In the full model, the fermions acquire a flavor index but any flavor symmetries are broken by the interaction term $S_\mathrm{IR}$~\eqref{sir}.  The Hamiltonian density is now $
H=H_0+H_{\rm int}$
where $H_0$ is a sum over a copy of~\eqref{h01} for each flavor and
\be
H_{\rm \spc int}=-\sum_{i_1,...i_q}J^-_{i_1...i_q}\psi^{i_1}_-...\, \psi^{i_q}_- \, -\sum_{j_1,...j_q}J^+_{i_1...i_q}  \psi^{j_1}_+...\,\psi^{j_q}_+.
\ee
Because $H_0$ is weakly vanishing we will drop it in what follows unless necessary. In isolation, the IR action is invariant under reparameterizations given the following modified transformation law for the fermions
\bea
\psi_\pm(z) \! & \mapsto &\! \Bigl|\det \frac{\p \tilde{z}^\mu}{\p z^\nu}\Bigr|
^{1/{q}}\psi'_\pm(\tilde{z}).
\eea
This is compatible with~\eqref{reparam} provided that
\bea
\Bigl|\det \frac{\p \tilde{z}^\mu}{\p z^\nu}\Bigr|\!\is\!1.
\eea
We thus see that the full theory is invariant under area-preserving diffeomorphisms. The charges generating these diffeomorphisms take the form
\be\label{contour}
W(\xi)=\oint dz^\mu \epsilon_{\mu\nu}\xi^\nu H_{\rm \spc int}, \qquad \qquad \p_\nu \xi^\nu =0
\ee
and obey a faithful representation of the lie algebra for the corresponding vector fields
\be
[W(\xi_1),W(\xi_2)]=-W([\xi_1,\xi_2]).
\ee
From our discussion in section~\ref{sec:sdg}, we recognize a $w_{1+\infty}$ symmetry in this model.

This appearance of a  $w_{1+\infty}$ symmetry, a conformal regime (to be discussed below), and maximal chaos are signs that the collective mode dynamics of the 2D SYK model shares important qualitative features with the celestial soft dynamics of 4D gravity. In the following we aim to make this similarity more explicit and present some initial entries of a potential holographic dictionary.

\subsubsection*{Bi-local collective field theory}

\vspace{-1mm}

At large $N$, the collective dynamics of the 2D SYK model is encoded in the fermionic two-point function and the collective frame field
\be
G_\pm(z_1,z_2)=\frac{1}{N}\sum_i \bigl\langle
\psi^i_\pm(z_1)\psi^i_\pm(z_2)\bigr\rangle,
\quad & &\quad \label{Nframe}
e^\pm=\frac{1}{N}\sum_i e^\pm_i.
\ee
While there are $N$ Cartan frames to integrate out, at leading order in large $N$ only this collective mode~\eqref{Nframe} contributes \cite{Turiaci:2017zwd}. The cubic kinetic term in~\eqref{s_uv2} makes it more convenient to go directly to the effective action in terms of bosonic bi-local dynamical mean fields, namely the two-point functions $G_\pm(x_1,x_2)$ and self energies $\Sigma^\pm(x_1,x_2)$.  Performing the disorder average and integrating out fermions, one obtains a bi-local dynamical mean field theory with the following effective action  
\bea\label{seff_bos}
S/N \! \is\!
-\int e^+\!\! \wedge \nspc e^- 
-\sum_{a=\pm}\log\mathrm{Pf}(e^a\!\!\wedge\nspc \partial -\Sigma^a)+\frac{1}{2}\int\!\!\int\!\Bigl(\Sigma^aG_a\!-\nspc\frac{J^2}{q}\bigl(G_+^q + G_-^q\bigr)\Bigr).
\eea
This bosonic effective action  shares the same symmetry under area-preserving diffeomorphisms as its fermionic counterpart~\eqref{fermi}. The resulting Schwinger-Dyson equations read
\be
\label{sdeqns}
\Sigma^\pm \!\! \is \!\! J^2 G_\pm^{q-1} ,  \qquad \ \ 
e^\pm\!\nspc\wedge\nspc \p G_\pm\! -\nspc 
\Sigma^\pm\!* G_\pm \spc = \, \delta_{12},\qquad \ \
e^\pm =\, \pm\spc {\p G_\mp} 
\big|_{z_2=z_1},
\ee
where $*$ denotes the convolution product and $\delta_{12}$ is a delta-function. Assuming translation invariance, the latter equation implies that the frame field $e^a$ is constant.

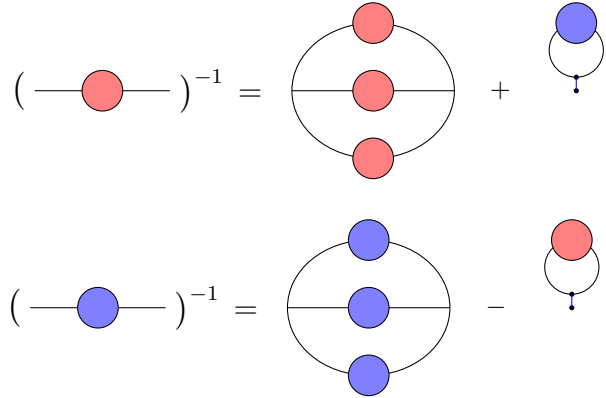
\begin{figure}[th!]
\begin{center}
\begin{tikzpicture}[scale=.9]
\draw (-1,0) node[left]{\mbox{\large$\bigl($}} -- (1,0) node[right]{\mbox{\large$\bigr)^{-1} \, = \  $}};
\draw[fill=red!50!white] (0,0) circle (3mm);
\draw[xscale=1.2] (3.333,0) circle (1cm);
\draw (2.8,0) -- (5.2,0) node[right]{\mbox{$\ \ \  +$}};
\draw[fill=red!50!white] (4,0) circle (3mm);
\draw[fill=red!50!white] (4,1) circle (3mm);
\draw[fill=red!50!white] (4,-1) circle (3mm);
\draw[xscale=1] (7,.6) circle (4mm);
\draw[fill=blue!50!white] (7,1) circle (3mm);
\draw[fill=black] (7,.2) circle (.3mm);
\draw[fill=black] (7,0) circle (.3mm);
\draw[blue] (7,0) -- (7,.2);
\end{tikzpicture}\\[5mm]
\begin{tikzpicture}[scale=.9]
\draw (-1,0) node[left]{\mbox{\large$\bigl($}} -- (1,0) node[right]{\mbox{\large$\bigr)^{-1} \, = \  $}};
\draw[fill=blue!50!white] (0,0) circle (3mm);
\draw[xscale=1.2] (3.333,0) circle (1cm);
\draw (2.8,0) -- (5.2,0) node[right]{\mbox{$\ \ \ -$}};
\draw[fill=blue!50!white] (4,0) circle (3mm);
\draw[fill=blue!50!white] (4,1) circle (3mm);
\draw[fill=blue!50!white] (4,-1) circle (3mm);
\draw[xscale=1] (7,.6) circle (4mm);
\draw[fill=red!50!white] (7,1) circle (3mm);
\draw[fill=black] (7,.2) circle (.3mm);
\draw[fill=black] (7,0) circle (.3mm);
\draw[blue] (7,0) -- (7,.2);
\end{tikzpicture}
\vspace{-4mm}
\end{center}
\caption{Diagrammatic representation of the Schwinger-Dyson  equations \eqref{sdeqns} with $q=4$.}
\end{figure}
\vspace{-2mm}

\def\tz{\zeta}
\def\bzz{{\bar \zeta}}
\def\ssigma{z}

\def\kkappa{\mbox{{\fontsize{11}{10}\selectfont  $\kappa$}}}

The bi-local fields live on the 4D kinematic space ${\cal K}_4$ parametrized by two pairs of coordinates $(z^\mu_1, z^\mu_2)$.  ${\cal K}_4$ is the symmetric product $\mathbb{R}^2 \times \mathbb{R}^2/\mathbb{Z}_2$ of two copies of 2D spacetime with a boundary given by the coincidence locus $z^\mu_1 = z^\mu_2$. Let us introduce lightcone coordinates on $\mathbb{K}_4$ via the projection along the frame-field\footnote{Here $\alpha =a =1,2$ in the first entry and $\dot \alpha = a-1 =0,1$ in the second.}
\bea
\label{czs}
(z_\aaa,\tz^{\dot \aaa}) \! \is \! (z^+_a, z^-_{a}) .
\eea  
We will use the $(z_\aaa,\tz^{\dot \aaa})$ notation whenever it reduces clutter. Kinematic space ${\cal K}_4$ comes equipped with a  metric with (2,2) signature defined by the bi-local vacuum expectation values 
\bea
\langle e^-(z_1) e^-(z_2) \rangle \, = \,\sqrt{\kkappa_+\!\!}\; \,
dz_1 dz_2,
\label{twovevs} 
  \quad & & \quad 
 \spc \langle e^+(z_1) e^+(z_2) \rangle\, =\, \sqrt{\kkappa_-\!\!}\;\,
d\tz^{\dot{0}} d\tz^{\dot{1}} 
\eea
with $\kkappa_\pm$ some constants. Here we built in that the frames by definition only have components in the corresponding chiral directions. We will use this (2,2) signature metric to raise and lower indices.

\subsubsection*{Large $q$ limit}\label{sec:doubles}
\vspace{-1mm}

Our goal in the following is to construct and study the effective field theory that describes the full soft dynamics of the 2D SYK model. A special limit that is particularly well suited for this purpose is the double scaling limit $q\to \infty, N\to \infty$ with $q^2/N$ fixed, where $q$ denotes the order of the SYK interaction. The effective action and equations of motion that describe the soft SYK dynamics considerably simplifies in this limit. As explored in more detail in the appendix, in the large $q$ limit the bi-local $(G,\Sigma)$ fields can be conveniently parametrized via an Ansatz~\cite{Maldacena:2016hyu,Cotler:2016fpe}
\bea\label{qansatz}
\Sigma^\pm \! \is \! \frac{\sigma_\pm(z_{1},z_2)}{q}, \qquad \quad G_\pm(z_1,z_2) \, =\, \frac{1}{2\sqrt{\kkappa_\mp}}\, {\rm sgn}(z_{12}^\pm) \left(1 + \frac{\pphi_\pm(z_1,z_2)}{q}\right)
\eea
where $\frac{1}{2\sqrt{\kappa_\mp}}{\rm sgn}(z_{12}^\pm)$ denotes the Green's function of the chiral fermions $\psi_\pm$ derived from the UV kinetic term \cite{Turiaci:2017zwd}.

It was shown in  \cite{Cotler:2016fpe} that the bi-local effective action of the 1D SYK model in the double scaling limit reduces to a Liouville action. Repeating the same analysis in the 2D case is not entirely straightforward due to the presence of the dynamical frame field. We will therefore adopt a mean field approximation and replace the frame fields $e^a$ and their bi-local products by the corresponding vacuum background values. With this extra physical input, we can follow the same calculation steps as in \cite{Cotler:2016fpe}. First we note that in the large $q$ limit, the bi-local SYK Hamiltonian $(G_+)^q+ (G_-)^q$ turns into the sum of two Liouville-type exponentials $e^{\pphi_+} + e^{\pphi_-}$. Next we expand the Pfaffian in \eqref{seff_bos} to second order in $\sigma_\pm$ and integrate out the $\sigma_\pm$ fields by performing the Gaussian  integral. This will produce a quadratic kinetic term for the $\pphi_\pm$ fields. The dynamical mean field two-forms \eqref{twovevs} provide the background metric for the sought after effective action.

Performing the calculation outlined above, we find that the effective action of the 2D SYK model in the double scaling limit takes the form of a sum of two Liouville actions
\bea
\label{biliouville}
S_{\rm eff} \!\! \is \!\! \frac{N}{4q^2} \int\!\nspc d^2z  d^2{\tz} \left(\nspc{\sqrt{\kkappa_-\!\!}}\;\spc \bigl( -  \frac 1 2 \p^\aaa\pphi_+ \nspc  \p_{\alpha} \pphi_+\!   +  {\cal J}^2 e^{\pphi_+}\bigr) + { \sqrt{\kkappa_+\!\!}}\; \spc \bigl(- \frac 1 2    \smpc \p^{\dot{\aaa}} {\pphi}_-\nspc \p_{\dot{\aaa}} {\pphi}_-\!      +  {\cal J}^2 e^{\pphi_-}\bigr)\right).
\eea 
The above effective theory looks like a higher dimensional version of the  effective theory of 1D double scaled SYK \cite{Cotler:2016fpe}. The 1D theory can be solved exactly by treating the Liouville interaction perturbatively \cite{DouglasTalk}. This leads to a perturbative expansion in terms of chord diagrams \cite{Berkooz:2018jqr} which can be summed exactly. We will not attempt to perform the same analysis here, but we will make temporary use of a similar perturbative treatment later on to exhibit the $w_{1+\infty}$ current algebra symmetry of 2D SYK theory.

The two individual terms in the action \eqref{biliouville} are symmetric under the product of 2D conformal transformations  and 2D area preserving diffeomorphisms, but each in a different combination. The first term is invariant under ($\zeta$-dependent) conformal transformations
in the $z$ plane 
\bea
(z_1,z_2)& \to & (Z_1(z_1,\zeta),
Z_2(z_2,\zeta)) 
\eea
provided $e^{\pphi_+}$ transforms as a $(1,1)$-form,
and ($z$-dependent) area preserving diffeomorphisms 
in the $\tz$ plane
\bea
\zeta^{\dot \aaa} &\to& F^{\dot \aaa}(\zeta,z), \qquad \qquad \det\bigl(\p_{\dot\aaa} F^{\dot\bbb}\bigr) = 1,
\eea
and vice versa for the second term. This special symmetry structure arises because the kinetic term of $\pphi_+$ and $\pphi_-$ only contain derivatives along two of the four directions of kinematic space $\mathbb{K}_4$.  \footnote{Normally, one might feel somewhat uncomfortable with field theories with ultra-local kinetic terms like in \eqref{biliouville}. They are acceptable, however, as effective field theories with a more complete UV description.}  

In the next section we will study the soft dynamics associated with these symmetries and make a  comparison with celestial holography.
For this application, we will mostly be interested in taking the strict large $N$ limit. In this limit, the effective dynamics of the bi-Liouville theory is restricted to the manifold of classical solutions to the equations of motion \bea
\label{eom}
\p^\aaa\p_\aaa\pphi_+ \! + {\cal J}^2 e^{\pphi_+} =\, 0, \quad & & \quad \; \, {\p}^{\dot \aaa}  {\p}_{\dot \aaa} \pphi_- \! + {\cal J}^2 e^{\pphi_-} =\, 0.
\eea
The effective action thus further simplifies to a sum of two JT gravity-like theories
\bea
\label{bijt}
S_{\rm eff}  \!\! \is \!\!  \int\!\nspc d^2z\, d^2\tz\, \left(  \sqrt{{\kkappa_-}\!\!}\; \,
\Phi^+ \bigl( \p^\aaa\p_{\aaa} \pphi_+ \nspc     \, +\,  {\cal J}^2 e^{\pphi_+} \bigr) \,+\,  \sqrt{\kkappa_+\!\!}\,\, \Phi^- \bigl(\p^{\dot{\aaa}}  \p_{\dot{\aaa}} {\pphi}_-\nspc  + {\cal J}^2 e^{\pphi_-} \bigr)\right)
\eea 
where the dilaton fields $\Phi^+$ and $\Phi^-$ act as Lagrange multipliers imposing the equations of motion \eqref{eom}. Note that both theories \eqref{biliouville} and \eqref{bijt} live in a (2,2) signature spacetime, even if the original SYK model is defined with Euclidean signature.

\medskip

\section{Soft Dynamics and Gravity}\label{sec:toy_dual}

The goal of this section is to merge our understanding of the symmetries of the 2D SYK model from section~\ref{sec:2DSYK} with the celestial symmetries reviewed in section~\ref{sec:celestial_w}. We will see that  the bi-local effective theory \eqref{biliouville} or \eqref{bijt} of the double scaled SYK model incorporates both Virasoro symmetry as well as the full $w_{1+\infty}$ current algebra symmetry. The Ward identities of these chiral currents comprise the full information about the gravitational MHV amplitudes \cite{Mason:2009afn}. On the geometric side, we will argue that the soft modes that distinguish the vacuum states of the 2D SYK model satisfy an equations of motion that looks identical to that of self-dual gravity.

\subsubsection*{\texorpdfstring{Virasoro symmetry}{w-infinity}}
 We start with a few comments about the Virasoro symmetry and soft dynamics. The bi-Liouville theory \eqref{biliouville} has a traceless  stress tensor in the $z$ plane 
 \bea
 T_{11} =\, 
 {-\frac{1}{2}}(\p_1\pphi_+)^2\! +\frac 1 2 \p_1^2\phi_+, \qquad T_{22}\! \is \!  
 {-\frac{1}{2}}(\p_2\pphi_+)^2\! +\frac 1 2 \p_2^2\pphi_+, \qquad
T_{12} = 0
 \eea
 that is chirally conserved $
 \p_2 T_{11} = \p_1 T_{22} \, =\,  0.$
 Upon quantization, the modes 
 \bea
 L_n \!\! \is \!  \oint\frac{dz_1}{2\pi i}\, z_1^{n+1} T_{11}(z_1), \qquad  \qquad T_{11}(z_1) = \int\! d^2\zeta\, \sqrt{\kappa_-\!\!\!}\;\;\, T_{11}(z_1,\zeta)
 \eea
 generate a left-moving Virasoro algebra, and similarly for the right-movers. The central charge of this algebra is imaginary, due to the fact that the 2D CFT is Lorentzian while the action has a real pre-coefficient. 
 
 In the quantum theory, the choice of vacuum will break conformal invariance to the finite dimensional subgroup of global conformal transformations
 \bea
 z_1 \to \frac{a z_1 + b}{cz_1 + b}.
 \eea
 This unbroken global conformal symmetry will later be identified with Lorentz symmetry.
 The $\pphi_+$ and $\pphi_-$ fields can be viewed as the Goldstone modes associated with this spontaneous symmetry breaking of conformal symmetry. To make this explicit, let us act with a general $\zeta$ and $z_2$ dependent diffeomorphism $z_1 \to Z_1(z,\zeta)$ on the $z_1$ coordinate. Using that the $\pphi_+$ field transforms under diffeomorphism~as  
\bea
\pphi_+(z_1,z_2,\zeta) \to \pphi_+(Z_1,z_2,\zeta) + \log(\p_1 Z_1)
\eea
we find that the transformed action \eqref{biliouville} produces the following effective action for the diffeomorphism field
\bea
S_{\rm eff}[Z] \is \frac{N}{8q^2} \int\! d^2z \, d^2\zeta\,  \frac{\bar\p Z\,}{\p Z}\left(\frac{\p^3 Z}{\p Z} - 2 \Bigl(\frac{\p^2 Z}{\p Z}\Bigr)^2\right)
\eea 
with $Z=Z_1$, $\p = \frac{\p}{\p{z_1}}$ and $\bar\p=\frac{\p}{\p z_2}$. The above action can be recognized as a generalization (because of the extra $\zeta$ integral) of the Alekseev-Shatashvili action, the geometric action of the Virasoro algebra \cite{Alekseev:1988ce,Cotler:2018zff}. The stress tensor of this geometric action takes the form of the Schwarzian derivative. Using this fact, one can show that in a combined large $N$ and high temperature limit, it reduces to Schwarzian quantum mechanics \cite{Mertens:2017mtv}. This Virasoro mode is responsible for the maximal Lyapunov behavior of the 2D SYK model.

\subsubsection*{\texorpdfstring{${w_{1+\infty}}$ current algebra}{w-infinity}}

\vspace{-1mm}

In section 2, we saw that 4D gravity exhibits $Lw_{1+\infty}$ loop group symmetry. Can we identify this symmetry in the effective theory of our 2D SYK model? In section 3, we introduced a set of $w_{1+\infty}$ charges in the UV description of the SYK model, defined as integrals of the microscopic Hamiltonian density times the vector field over a spatial slice. In the bi-local disorder averaged theory, we expect that these global charges $\hat{w}^p_m$ are represented by a 3D surface integral of the time component of a conserved current. Moreover, one would expect that this current splits into two chirally conserved currents in the conformal regime.  To test this expectation we will now explicitly construct this current for the 4D JT gravity-like theory \eqref{bijt}.

Encouraged by the success of the perturbative treatment of the 2D effective theory of the 1D SYK model, we will assume that the 4D theory can be built up via a similar perturbative expansion. We write the 4D JT theory \eqref{bijt} as a sum 
\bea
S_{\rm eff} \!\is \! S_0 + \tilde{S}_0 + S_{\rm int}
\eea 
of two free boson actions and a Liouville interaction term. We will first analyze the symmetry structure of the free theory and then reintroduce the interactions. 
 
The non-interacting theory splits into a sum of two free boson actions
\bea
\label{freeboson}
S_{0}\!\! \is \!\! 
\int\! d^2z\, d^2 \tz  \sqrt{{\kappa_-}}\;\Phi^+\spc \p^\aaa\p_\aaa \pphi_+, 
\eea
plus an analogous kinetic term for ${\Phi}^-$ and ${\pphi}_-$. Let us focus on the above sector. From now on we will drop the $\pm$ sub/superscript. The free equation of motion $\p^\aaa\p_\aaa \pphi = \p_2\p_1 \pphi=0$ is solved by writing $\phi$ as a sum of a left and right-moving wave
\bea
\qquad \pphi(z,\bz,\zeta) \! \is \! \varphi(z,\zeta) + \bar{\varphi}(\bz,\zeta), \qquad \qquad\  z \equiv z_1, \quad \bz \equiv z_2.\qquad
\eea
In the quantum theory, $\varphi(z,\zeta)$ and $\bar{\varphi}(\bar{z},\zeta)$ represent two chiral bosons. To make the chiral factorization explicit, we introduce the two conjugate field $\pi = \p \Phi$ and $\bar\pi = \bar\p \Phi$ and recast the free boson action \eqref{freeboson} as
\bea
\label{chiralboson}
S'_{0} \!\! \is \!\!-\int \!\nspc d^2 z\spc d^2\tz \,  \textstyle \sqrt{{\kappa}}\,\bigl( \pi \bar\p \varphi + \bar\pi \p \bar\varphi\bigr) , \qquad \quad \  \overline\p \equiv \p_2, \quad \p \equiv \p_1.
\eea  
The equations of motion are $\bar\p\varphi = \p \bar\varphi =0$. This chiral boson action has the two symmetry groups that characterize celestial dynamics: it defines a 2D CFT with Virasoro symmetry in the $z$-plane and it has chiral area preserving diffeomorphism invariance in the $\tz$-plane.

Area preserving diffeomorphism invariance acts like an internal symmetry from the point of view of the 2D CFT. It acts on the $\varphi$ and $\pi$  fields via
\bea
\delta_v \phi \! \is \! \frac{1}{\sqrt{{\kappa}}}\;
\epsilon^{a \dot \bbb} \, \p_{\dot \aaa} v\,  \p\raisebox{1pt}{${}_{\dot \bbb}$}\phi, \qquad \qquad 
\delta_v \pi \, =\,  \frac{1}{\sqrt{{\kappa}}}\;
\epsilon^{a \dot \bbb} \, \p_{\dot \aaa} v\,  \p\raisebox{1pt}{${}_{\dot \bbb}$}\pi
\eea
where $v(\zeta)$ denotes an arbitrary smooth function of the transverse coordinates $\tz^{\dot \aaa}$. 
As before, let $v_{n}^p(\tz)$ denote the polynomial basis \eqref{wmp} of such functions. The corresponding infinite set of area preserving transformations are generated by the $w_{1+\infty}$ charges
\bea\label{winfq}
\ww^p_n\!\! \is \!\!  \oint\!\frac{d z}{2\pi i}\,  W^{p}_{n}(z),\ \qquad \ \qquad 
W^p_{n}(z) \spc = \spc  \int\! d^2 \tz \, v^p_n(\tz) \, T_1 (z,\tz)
\eea
where $T_1(z,\tz)$ denotes the chiral current
\bea\label{T1chiral}
T_{1}(z,\tz) \!\is \! \epsilon^{\dot{\alpha}\dot{\beta}} \p_{\dot \alpha} T_1\spc\spc\raisebox{1pt}{${\!}_{\dot\beta}$} \, = \,  \epsilon^{\dot{\alpha}\dot{\beta}} \p_{\dot \alpha}\pi\, \p\raisebox{1pt}{${}_{\dot\beta}$}\varphi 
\eea
with $T_1\spc\raisebox{1pt}{${\!}_{\dot\beta}$}$ the mixed components of the stress energy tensor.
Thanks to the fact that $\pi$ and $\varphi$ are chiral fields, this local currents $T_{1}(z,\tz)$ and $W^p_n(z)$ satisfy the chiral conservation law 
\bea
\bar\p \spc T_{1}(z,\tz)\! \is \! 0, \qquad \qquad \bar\p \spc W^p_n(z)\, = \, 0.
\eea
This allows the introduction of an infinite set of charges
\bea
\ww{\spc}^p_{n,r} \!\is\! \oint\! \frac{dz}{2\pi i}\, z^{r}\, W^p_n(z) 
\eea
that generate the $w_{1+\infty}$ current algebra. 

The above construction of chiral currents can in principle be extended to the interacting theory. This can be done in different ways. One way is to incorporate the interaction term order by order and show that the currents can be corrected so they remain chirally conserved at each order in perturbation theory. A more direct argument, that indicates that this can indeed be done, is  that each of the Liouville equations of motion \eqref{eom} can in fact be mapped to a free field wave equation
 via a non-linear B\"acklund transformation
\bea
\p\phi_0 \! \is \!  \p \phi + {\cal J} \spc e^{(\phi + \phi_0)/2}\, , \qquad \qquad
\bar\p\phi_0 \, = \, -\bar\p \phi  + {\cal J} \spc e^{(\phi -\phi_0)/2}\, .
\eea
If $\phi_0$ satisfies the linear wave equation $\p \bar\p \phi_0\! =\! 0$, then $\phi$ satisfies the non-linear Liouville equation \eqref{eom}, and vice versa. Although the field redefinition itself looks non-local, the free field $\phi_0$ still defines a local operator in the interacting theory. Moreover, since it is free, we can write it as a sum $\phi_0 = \varphi_0 + \bar\varphi_0$ of two chiral free fields with associated conjugate free fields $\pi_0$ and $\bar{\pi}_0$. The chiral currents of the interacting theory are then constructed in terms of the free fields $\varphi_0$ and $\pi_0$. The explicit expression of the chiral currents in terms of the interacting fields looks non-local, but the currents themselves act as local operators. In the following we will continue to work with the free field expressions, but we will drop the subscript${}_0$.

The emergence of a $w_{1+\infty}$ current algebra is associated with the fact that the vacuum state in the conformal regime spontaneously breaks the area preserving symmetry. We will now summarize how this soft symmetry breaking and the Ward identities of the chiral currents comprise the full information about gravitational MHV amplitudes and self-dual solutions to 4D Einstein gravity.

\subsubsection*{Soft modes and self-dual 4D gravity}

\vspace{-1mm}

 Let ${h}(z,\tz)$ be a general (0,1)-form valued function of $z$ and $\tz^{\dot \alpha}$.  For now we take $h$ infinitesimally small. We can deform the free action  \eqref{chiralboson} via
\bea
\label{deformaction}
S'_0 + \int\!\! dz \spc d^2\nspc \zeta\spc {\textstyle{\sqrt{{\kappa}}}}\, h\spc T_{1}  \!\is \! -\int \!\! d z \spc d^2\nspc \zeta\spc {\textstyle{\sqrt{{\kappa}}}} \,  \pi\bar{\nabla}\varphi
\eea
where $\bar{\nabla}$ denotes the deformed  Dolbeault operator
\bea\label{db}
\bar{\nabla}\! \is\!  \bar{\p}+\epsilon^{\dot \aaa\dot \bbb}\frac{\p{h}}{\p\tz^{\dot \aaa}}\frac{\p}{\p\tz^{\dot \bbb}}  \equiv \bar{\p}+\{\spc h, \ \, \}.
\eea
with $\bar\p = d\bar{z} \frac{\p}{\p \bar{z}}$.
The deformed action can be viewed as the result of applying an $z$ dependent area preserving diffeomorphism to the vacuum state of the original undeformed free theory. Let us make this interpretation explicit and simultaneously generalize to the case that $h$ represents a finite deformation.

Starting from the standard $\mathrm{SL}(2,\mathbb{R})$ invariant vacuum $|0\rangle$, we introduce a new vacuum 
\bea
|\spc 0 \spc \rangle_{{h}} \!\is \!  U(F)|\spc 0\spc \rangle
\eea
with a non-trivial $w_{1+\infty}$ soft mode by acting with a unitary $U(F)$ that implements $z$ dependent area preserving diffeomorphism $
F :\, (z,\tz^{\dot \aaa}) \mapsto  \bigl(z,F^{\dot \aaa}(z,\tz)\bigr),$ with $F$ is related to the (0,1)-form valued function~${h}$~via 
\bea
\label{fhrel}
\bar\partial F^{\dot \aaa} \!\! \is  \! \epsilon^{\dot \aaa\dot \bbb}\frac{\p  {h}}{\p \tz^{\dot \bbb}}. \eea
In the quantum theory, this relation implies that $[\bar{L}_{-1}, U(F)] = \ww(h)U(F)$, where $\ww({h})$ denotes the Noether charge that generates the linear transformation $\bigl[\hat{w}({h}),{\cal O} \bigr] = 
\bigl\{{h}, {\cal O}\bigr\}$ on local operators~${\cal O}$.
The transformed vacuum $|\spc 0 \spc\rangle$ is not translation invariant, but satisfies 
\bea
\bar{L}_{-1} |\spc 0 \spc \rangle_{{h}}\! \is \! \ww(h) |\spc 0 \spc \rangle_{{h}}.
\eea
It is instructive to relate the (0,1)-form valued function $h$ to the expectation value of the chiral field $\varphi(z,\zeta)$ in the deformed vacuum. In the standard vacuum $
\langle \spc \bar\p \varphi \rangle  = \langle \spc 0\spc| \,[\bar{L}_{-1}, \varphi]\, |\spc 0\spc\rangle \, = \, 0$, implying that
the expectation value of $\varphi$ is a holomorphic function in $z$. In the deformed vacuum we instead find that $
\langle \spc \bar\p \varphi  \rangle = \langle \spc [\bar{L}_{-1}, \varphi] \spc\rangle
=\langle \spc [\ww({h}), \varphi ]\spc \rangle\, = \langle\spc \{{h},  \spc \varphi\}\rangle.$
Hence we see that $\langle\varphi\rangle$ is holomorphic with respect to the deformed Dolbeault operator \eqref{db}, $
\bar{\nabla} \langle  \varphi \rangle = 0.$ This matches the equation of motion of the deformed action \eqref{deformaction} in the linearized regime.
We can generalize this reasoning to show that the deformed vacuum expectation values of all holomorphic currents and conformal blocks are annihilated by $\bar\nabla$.\footnote{The integrability requirement $\bar\nabla^2 =0$ implies $\bar\p h + \frac 1 2 \{h,h\} = 0$; in 2D this is automatically satisfied \cite{Adamo:2021bej}.}

The above discussion closely follows the non-linear graviton description of classical self-dual geometries in 4D gravity \cite{Penrose:1976js}. To make this relationship explicit we need to lift the story to 4D spacetime via the twistor correspondence \cite{Adamo:2021bej}. 
We define twistor variables $(\lambda_\aaa,\zeta^{\dot  \alpha})$ via
\bea\label{curve1}
\lambda_{\aaa} \, = \, \bigl(\spc1\spc,\spc z\bigr) ,~ & & ~\mmu^{\dot \aaa}\,=\,  {\tz^{\dot \aaa}}. 
\eea
A given point $x^{\alpha \dot \alpha}=(x^{\dot \aaa},\tilde{x}^{\dot \alpha})$ in flat Minkowski spacetime specifies a linear curve in twistor space
\bea    
\label{tline}
\zeta^{\dot \aaa} \! \is \! x^{\dot{\aaa} \alpha}\lambda_\aaa 
\, =\,  {x^{\dot \aaa}}
+ {\tilde{x}^{\dot \aaa}} \smpc z .
\eea 
The non-linear graviton construction of Penrose considers deformations of the complex structure on twistor space  $\mathbb{PT}$. These deformations amount to turning on the soft $w_{1+\infty}$ mode and can be implicitly parametrized by replacing the linear relation \eqref{tline} by a general degree one curve 
\be\label{curve1}
\zeta^{\dot \aaa} \! \is\!  {x^{\dot \aaa}}+  {\tilde{x}^{\dot \aaa}}\spc z +  Z^{\dot \aaa}(\zeta,z)
\ee
where $Z^{\dot \aaa}$ is a smooth function in $z$. The relation $\bar\p  Z^{\dot \aaa} = \epsilon^{\dot \aaa\dot \bbb} \frac{\p  {h}(x,z)}{\p \mmu^{\dot \bbb}}$
guarantees that the curve~\eqref{curve1} is holomorphic with respect to the $\bar{\nabla}$ operator~\eqref{db}. 

As shown in  \cite{Adamo:2021bej}, one can write down a sigma model which produces this relation between $Z^{\dot\alpha}$ and $h$ as its equation of motion
\be
S[Z]=
\int_{\mathbb{CP}^1}\!\!\! dz\, \bigl(\epsilon_{\dot \aaa\dot \bbb} Z^{\dot \aaa}\bar\p Z^{\dot \bbb}+2{h}(z, Z)\bigr)
\ee
where $Z^{\dot \aaa}$ is treated as a function of $x^{\aaa\dot \aaa}$ and $z_\aaa=(1,z)$.
Upon minimizing the action with respect to the dynamical field $Z^{\dot \aaa}(x,z)$, and after integrating over $\mathbb{CP}^1$, the only remaining dependence is on the spacetime point $x$. The tetrad of the corresponding self-dual geometry is then obtained via \cite{Adamo:2021bej}
\be\label{omtetrad}
e^{a\dot \aaa}=(dx^{\dot \aaa},\Omega^{\dot \aaa}_{~\dot\beta}d\tilde{x}^{\dot \aaa}),~~~\Omega_{\dot \aaa\dot \aaa}=\frac{\p^2\Omega}{\p x^{\dot \aaa}\p \tilde{x}^{\dot \bbb}}, \qquad \ \Omega(x)=\epsilon_{\dot \aaa\dot \bbb}x^{\dot \bbb}\tilde{x}^{\dot \aaa}\! - \nspc S[F]\big|_{\rm on-shell}.
\ee
The usefulness of introducing the twistor sigma model is that the on-shell action is a generating functional for MHV amplitudes in 4D gravity, where the $x$ and  $\tilde{x}$ are (half-)Fourier transformed to the spinor helicity variables for the two negative helicity particles~\cite{Mason:2009afn}. 

\subsubsection*{Graviton vertex operators}

The conformal primary wavefunctions with positive helicity are examples of self-dual Ricci-flat spacetimes, pertinent to the non-linear graviton construction~\cite{Pasterski:2020pdk, Adamo:2021bej}.  Let us use this to make the connection between the area preserving diffeomorphisms of our model and the celestial $w_{1+\infty}$ symmetry explicit. We would like to write a dictionary between 

\addtolength{\parskip}{-1mm}
\begin{itemize}
\addtolength{\baselineskip}{-2.5mm}
\addtolength{\parskip}{-1mm}
\item{the chiral currents $ T_1(z, \tz)$ and $W^p_n(z)$ of our 2D SYK model~\eqref{T1chiral}}
\item{the momentum space graviton operators 
 $\widetilde{H}(z,\tilde\lambda)$ with momentum  $p_{\aaa\dot \aaa} = \lambda_\aaa \tilde{\lambda}_{\dot \aaa}$ with $\lambda_\aaa = (1,z){{\strut}}$}
 \item{the celestial operators $H^{-2p+4}(z,\biz)$ in \eqref{hk} that incorporate the $w_{1+\infty}$ currents in celestial CFT}
 \item{the spacetime graviton field
$H^{-2p+4}_{\alpha\beta\dot\alpha\dot\beta}(x)$ associated with the celestial operators $H^{-2p+4}(z,\tilde{z})$}
\addtolength{\parskip}{1mm}
\addtolength{\baselineskip}{2.5mm}
\end{itemize}
\addtolength{\parskip}{1mm}

We propose that the momentum operators and the current $T_1(z, \tz)$ are related via a Fourier transform
\bea
\widetilde{H}(z,\tilde\lambda) \!\is \!
 \int\!d^2\zeta\,  T_1(z, \tz) \, e^{-i\zeta \cdot \tilde{\lambda}}\, .
\eea
The special celestial operators are then given by applying the Mellin transform
\bea
\label{wpnzz}
H^{-2p+4}(z,\biz) 
\! \is \! \lim_{\epsilon\to 0}\; \epsilon  \int \! d\omega \, \omega^{-2p+1+\epsilon\, } \, \widetilde{H}(z,\omega\tilde{z})  \\[2mm]
\is \lim_{\epsilon\to 0}\; \epsilon  \int \! d\omega \, \omega^{-2p+1+\epsilon\, } \!\int\!d^2\zeta\,  T_1(z, \tz) \, e^{-i\omega\zeta \cdot \biz}
\label{htone}
\eea
where $\biz_{\dot \aaa} = (1,\biz)$.  We see that the mapping from the kinematic space of the 2D SYK model to the celestial sphere is not direct, but involves a chiral projection on the $z_1$-direction and a Fourier plus Mellin transform.

 We can verify this proposal by matching the $Lw_{1+\infty}$ generators in~\eqref{hk} to those in~\eqref{winfq}.  Doing the Mellin integral in \eqref{htone} first, we have
\be\label{mellin}
h^{p}(\zeta,\biz) \, \equiv\, \lim_{\epsilon\to 0}\; \epsilon  \int \! d\omega \, \omega^{-2p+1+\epsilon} 
    e^{-i\omega \tz \cdot \tilde{z}}
\!\is \!  \; \frac{
 (\tz \cdot \tilde{z})^{2p-2}\!\!\!\!\!}{\Gamma(2p-1)}.
\ee
Expanding this in $\biz$ we get
 \be
h^{p}(\zeta,\biz)=\sum_n \frac{\tilde{z}^{p-n-1}}{\Gamma(p-n)\Gamma(p+n)}\, v^{p}_n(\zeta)
\ee
so that, indeed
\bea
\label{wpnzz2}
H^{-2p+4}(z,\tilde{z}) \! \is \! \int\!d^2\zeta\, \,  h^{p}(\zeta,\biz) T_1(z,\zeta)=\sum_n \frac{\tilde{z}^{p-n-1}}{\Gamma(p-n)\Gamma(p+n)}\, W^p_n(z)\, 
\eea
and the $Lw_{1+\infty}$ generators of the two models match. Comparing order by order in~$\tilde{z}$, we see  that the celestial CFT vertex operators associated with the $w_{1+\infty}$ soft modes with given momentum are obtained  by multiplying the conserved chiral current $T_1(z, \tz)$ by the corresponding $w_{1+\infty}$ mode function $v^p_n(\tz)$, matching~\eqref{winfq}.

The integral kernel relating the celestial primaries and chiral SYK current in~\eqref{wpnzz2} is precisely an uplift of the scalar conformal primary wavefunction to twistor space with
\bea
h^{p}(x,z,\biz) = \frac{
 (-q\cdot x)^{2p-2}\!\!\!\!\!}{\Gamma(2p-1)}\! \is \! h^{p}(\zeta,\biz)\bigr|_{\zeta^{\dot\aaa} = x^{\dot\aaa \aaa}\lambda_\aaa}
\eea
on the incidence relation 
with  $\lambda_\alpha = (1,z)$. Indeed, using the twistor correspondence we can make the holographic dictionary between our chiral currents and the spacetime gravitons explicit for the full graviton modes, including the dependence on polarization tensors, as follows
\bea
H_{\alpha\beta\dot\alpha\dot\beta}^{-2p+4}(x)
\!\is\! \iota_\alpha\iota_\beta
\int\! d^2 z  \,  h^{p-1}_{\dot\alpha\dot\beta}(x,z,\biz)
\, H^{-2p+4}(z,\tilde{z})
\eea
with $\iota_\alpha=(0,1)$. Here $\int d^2z=\oint\frac{dz}{2\pi i} \oint\frac{d\tilde{z}}{2\pi i}$  is the analytic continuation of the celestial sphere measure to the celestial torus\footnote{In each of these expressions we want to formally analytically continue $p$ away from integer values to do the Mellin integral.  Upon performing the shift $p\rightarrow p_\epsilon=p-\frac{1}{2}\epsilon'$ then the residues
$
\lim_{\epsilon'\rightarrow0}\epsilon'h_{\alpha\beta\dot\alpha\dot\beta}^{p_\epsilon}(x)$
are finite.  One would precisely get a sum over such residues for each $p$ in $p=1,\frac{3}{2},2,\frac{5}{2},...$ if we started from a basis expansion of the graviton operator on radiative states in the self dual (positive helicity) sector and deformed the principal series contour to the left, picking up the poles that appear at the location of the $Lw_{1+\infty}$ generators.} and 
\bea
h^{p-1}_{\dot\alpha\dot\beta}(x,z,\biz) \!\is \!
\frac{\p^2}{\!\p \zeta^{\dot\alpha} \p \zeta^{\dot\beta}\!}\;
h^{p}(\zeta,\tilde{z})\bigr|_{\zeta^{\dot\aaa} = x^{\dot\aaa \aaa}\lambda_\aaa}.
\eea

From this discussion we see that the chiral SYK current is closer to the twistor transformed data for the positive helicity external states. Explicitly, inverse Mellin transforming~\eqref{wpnzz} leaves us with the half-Fourier transform to the twistor basis for scattering~\cite{Witten:2003nn,Arkani-Hamed:2009hub}. Composing Mellin and twistor transforms has a natural interpretation in terms of light transforms of the celestial operators~\cite{Sharma:2021gcz}.

\section{Conclusions}\label{sec:conclusions}

\vspace{-1mm}

In this paper we have proposed that a variant of the 2D SYK model introduced in~\cite{Turiaci:2017zwd} provides a toy model for the soft limit of the gravitational sector in 4D asymptotically flat spacetimes. We have motivated this by matching the symmetries on both sides as well as the currents and soft mode dynamics. The effective dynamics of the Virasoro soft modes of the 2D SYK model matches with the superrotation soft dynamics in celestial CFT.   Moreover, we have shown that the area preserving diffeomorphism symmetry of this model match onto the $w_{1+\infty}$ symmetry acting on the self-dual sector of 4D gravity. 

In combination, these results hint at a structure that ties together the observation of our recent paper \cite{Pasterski:2022lsl} on how signals of maximal quantum chaos are encoded in the superrotation Goldstone modes with the presence of celestial $w_{1+\infty}$ symmetry~\cite{Strominger:2021lvk}. Fortuitously, while the perspective in~\cite{Pasterski:2022lsl} ignored the $w_{1+\infty}$ structure, it naturally led us to revisit 2D SYK models of the type proposed in~\cite{Turiaci:2017zwd}.  These models precisely have these extra symmetries but hitherto lacked an obvious application or purpose for them. Here we have presented evidence that the parallel between the dynamics of the soft sectors of 4D gravity and the 2D SYK model is more than just a superficial similarity. Much more work needs to be done, however, to find out if the application of the 2D SYK model to celestial soft dynamics can be lifted into a well stated holographic duality or not.

The emerging narrative ends up taking on some longstanding questions about celestial holography. Is the dual theory on the celestial sphere a genuine 2D CFT? Can we identify specific 2D quantum field theories or quantum many body systems that have the same soft dynamics and low energy symmetries as 4D gravity? How does celestial holography relate to AdS holography?  In each case the answer requires a careful understanding of the vacuum structure of asymptotically flat spacetimes, its symmetries, and the role of IR regulators.  While we have focused on the soft limit of gravitational sector, we have seen that there is plenty of interesting physics governed by these symmetries.   Going beyond the soft sector, we expect to gain more insight into the possible microscopic realizations of celestial holography, while the same overabundance of symmetries that make the celestial holographic dual exotic as a CFT will give us more power to constrain and compute gravitational  scattering amplitudes.

\subsection*{Acknowledgements}
\vspace{-1mm}

We thank Laurent Freidel, Akash Goel, Andrew Strominger, Lee Smolin, Joaquin Turiaci, and Erik Verlinde for valuable discussions and comments.  The research of SP is supported by the Sam B. Treiman Fellowship at the Princeton Center for Theoretical Science.  The research of HV is supported by NSF grant number PHY-1914860.

\appendix
 
 \section{From Schwinger-Dyson to Liouville}

 In this appendix we show how the Schwinger-Dyson equations for our 2D SYK model land us on the Liouville equation. Here we restrict to translation invariant solutions with the property that the bi-local fields only depend on the coordinate difference between the two arguments. The third SD equation in \eqref{sdeqns} then implies that the background frame field is constant. The remaining SD equations subsequently factorize into two decoupled chiral set of equations. This will enable us to follow a generalization of the derivation in~\cite{Maldacena:2016hyu} for our two chiral copies, so long as we can motivate there are no divergences associated with the presence of an extra direction. The derivation below is meant to elucidate the assumptions and range of validity of the effective field theory. 
 
Following \cite{Maldacena:2016hyu}, we start with the Ansatz  
\bea
\label{msansatz}
G_\pm(z)\! \is \! \frac{1}{2\sqrt{\kkappa}}\,\spc \mathrm{sgn}(z^\pm)\Bigl( 1 + 
\frac 1 q {\pphi_\pm(z)}\Bigr)\, , \qquad \qquad \Sigma^\pm(z) \, = \, \frac{\sqrt{\kkappa}\, {\cal J}^2 }
{q } \mathrm{sgn}(z^\pm) e^{\pphi_\pm(z)} 
\eea 
which solves the large $q$ limit of the first SD equation $\Sigma^\pm(z) =  J^2 G_\pm^{q-1}(z)$  provided we set ${\cal J}^2 =\frac {2q J^2}{(2\sqrt{\kappa})^{q}}$. Here we set $\kappa_\pm = \kappa$, and $z = (z^+,z^-) = (z^+_{12},z^-_{12})$ denotes the coordinate difference.
The second SD equation in \eqref{sdeqns} is local in frequency space $\omega = (\omega_+,\omega_-)$ and takes the form
\bea
\label{sdtwo}
\frac{1}{G_\pm(\omega)} \! \is \! -{i\sqrt{\kappa}}\, \omega_\pm\; -\spc \Sigma^{\pm}(\omega).
\eea
Here we used our assumption that the frame field is constant.
Plugging the Ansatz \eqref{msansatz} for $G_\pm$ gives that 
\bea
\label{mstwo}
\frac{1}{\! G_\pm(\omega)\!}  \is \! \frac{ \sqrt{\kappa}}{- \frac{1}{i\omega_\pm} \spc + \frac{1}{2q}[\mathrm{sgn}(z^\pm)\pphi_\pm(z)](\omega)}
= -{i \sqrt{\kappa}\, \omega_\pm}\spc
+\spc  \frac{ \sqrt{\kappa}\,  \omega^2_\pm }{2q} \, [\mathrm{sgn}(z^\pm)\pphi_\pm(z)](\omega) + \ldots
\eea
where ... denote terms that vanish in the large $q$ limit. To justify the second equality, we need to assume~that 
\bea
\label{irbound}
[{\rm sgn}(z^\pm)\pphi_\pm(z)](\omega) \; \ll \; \frac{2 q } {\omega_\pm }
\eea 
for all values of $\omega_+$ and $\omega_-$. Comparing equations \eqref{mstwo} and \eqref{sdtwo}, one derives that
\bea
2\p_\pm^2\bigl(\mathrm{sgn}(z^\pm) \pphi_\pm(z)\bigr) \! \is \!  {\cal J}^2\, \mathrm{sgn}(z_\pm) e^{\phi_\pm(z)}
\eea
which implies the two Liouville equations of motion 
\bea
\label{eomtwo}
2\p_1\p_2\pphi_+ \! + {\cal J}^2 e^{\pphi_+} = \, 0, \quad & & \quad \; \, 
2\bar{\p}_{\dot 2}  \bar{\p}_{\dot 1} \pphi_- \! + {\cal J}^2 e^{\pphi_-}=\, 0
\eea
away from the boundary $z_{12}^\pm = 0$ of kinematic space. 
Here $\p_a\! =\! \frac{\p}{\p z^+_a}$ and $\bar\p_{\dot a} \! =\! \frac{\p}{\p z^-_a}$. 

The two Liouville equations of motion \eqref{eomtwo} have the following general solution
\bea
{\cal J}^2 e^{\phi_+} = \, \frac{4\p_1 Z^+_1 \p_2 Z^+_2}{ (Z^+_1\! -Z^+_2)^2}, \quad & & \quad {\cal J}^2 e^{\phi_-} = \, \frac{4\bar\p_{\dot 1} Z^-_1 \bar\p_{\dot 2} Z^-_2}{ (Z^-_1\! -Z^-_2)^2}
\eea 
where $Z^\pm_a(z_a)$ are arbitrary functions of the corresponding coordinates~$z_a^\pm$.
However,  the range of validity of these solutions is somewhat restricted. First,  in the derivation of the effective SD equations \eqref{eomtwo},
we made the explicit assumption that the frame field $e^\pm$ is constant and that the bi-local fields only depend on the coordinate difference. It seems reasonable, however, that these conditions can be somewhat relaxed to the adiabatic assumption that the spatial variations of the frame metric and of the bi-local fields on the center of mass coordinate varies very slowly compared to their spatial dependence on the relative coordinate distance.


\bibliographystyle{utphys}
\bibliography{cCFT}

\end{document}